\documentclass[12pt]{article}

\usepackage{scicite}

\usepackage{times}
\usepackage{amsmath}
\usepackage{amsfonts}
\usepackage{mathtools}
\usepackage{graphicx}
\usepackage[]{color}
\newcommand{\bl}[1]{\color{black}{#1}}

\newcounter{lastnote}

\title{Nonlinear amplitude dynamics in flagellar beating}

\author
{David Oriola,$^{1,\ast}$ Hermes Gad\^elha,$^{2,3}$ Jaume Casademunt $^{1}$\\
\\
\normalsize{$^{1}$Departament de F\'isica de la Mat\`eria Condensada} \\
\normalsize{Facultat de F\'\i sica, Universitat de Barcelona} \\
\normalsize{Avinguda Diagonal 647, E-08028 Barcelona, Spain.}\\
\normalsize{$^{2}$Department of Mathematics, University of York, YO10 5DD, UK.}\\
\normalsize{$^{3}$Wolfson Centre for Mathematical Biology, Mathematical Institute} \\
\normalsize{University of Oxford, Oxford OX2 6GG, UK. } \\
\\
\normalsize{*Current address:} \\
\normalsize{Max Planck Institute of Molecular Cell Biology and Genetics} \\
\normalsize{Pfotenhauerstra{\ss}e 108, 01307} \\
\normalsize{\&} \\ 
\normalsize{Max Planck Institute for the Physics of Complex Systems} \\
\normalsize{N\"othnitzerstra{\ss}e 38, 01187} \\
\normalsize{Dresden, Germany } \\
\\
\normalsize{To whom correspondence should be addressed; E-mail:  oriola@mpi-cbg.de, oriola@pks.mpg.de.}
}

\date{}

\begin{document} 


\baselineskip24pt


\maketitle 

\clearpage




\section*{Abstract}
The physical basis of flagellar and ciliary beating is a major problem in biology which is still far from completely understood. The fundamental cytoskeleton structure of cilia and flagella is the axoneme, a cylindrical array of microtubule doublets connected by passive crosslinkers and dynein motor proteins. The complex interplay of these elements leads to the generation of self-organized bending waves. Although many mathematical models have been proposed to understand this process, few attempts have been made to assess the role of dyneins on the nonlinear nature of the axoneme. Here, we investigate the nonlinear dynamics of flagella by considering an axonemal sliding control mechanism for dynein activity. This approach unveils the nonlinear selection of the oscillation amplitudes, which are typically either missed or prescribed in mathematical models. The explicit set of nonlinear equations are derived and solved numerically. Our analysis reveals the spatiotemporal dynamics of dynein populations and flagellum shape for different regimes of motor activity, medium viscosity and flagellum elasticity. Unstable modes saturate via the coupling of dynein kinetics and flagellum shape without the need of invoking a nonlinear axonemal response. Hence, our work reveals a novel mechanism for the saturation of unstable modes in axonemal beating.  \\

{\bf Keywords}: Flagellar beating, Dynein, Spermatozoa, Self-organization.

\clearpage



\section{Introduction}
Cilia and flagella play a crucial role in the survival, development, cell feeding and reproduction of microorganisms \cite{Ginger}. These lash-like appendages follow regular beating patterns which enable cell swimming in inertialess fluids \cite{Lauga}. Bending deformations of the flagellum are driven by the collective action of ATP-powered dynein motor proteins, which generate sliding forces within the flagellar cytoskeleton, named axoneme \cite{Satir_68}. This structure has a characteristic `9+2' composition across several eukaryotic organisms, corresponding to 9 peripheral microtubule doublets in a cylindrical arrangement surrounding a central pair of microtubules \cite{Ginger}. Additional proteins, such as the radial spokes and nexin crosslinkers, connect the central to the peripheral microtubules and resist free sliding between the microtubule doublets, respectively. Each doublet consists of an A-microtubule in which dyneins are anchored at regular intervals along the length of the doublets, and a B-microtubule, where dynein heads bind in the neighbouring doublet \cite{Gaffney_review,Ginger}. In the presence of ATP, dyneins drive the sliding of neighbouring microtubule doublets, generating forces that can slide doublets apart if crosslinkers are removed \cite{Summers}. In the presence of crosslinkers, sliding is transformed into bending. Remarkably, this process seems to be carried out in a highly coordinated manner, in such a way that when one team of dyneins in the axoneme is active, the other team remains inactive \cite{Mitchison}.  This mechanism leads to the propagation of bending undulations along the flagellum, as commonly observed during the movement of spermatozoa \cite{Gaffney_review}. \\

Many questions still remain unanswered on how dynein-driven sliding causes the oscillatory bending of cilia and flagella \cite{Brokaw_review}. Over the last half a century, intensive experimental and theoretical work has been done to understand the underlying mechanisms of dynein coordination in axonemal beating. Different mathematical models have been proposed to explain how sliding forces shape the flagellar beat \cite{Machin, Brokaw_71, Hines-Blum, Hines-Blum2, Lindemann, Camalet, Camalet_NJP, Riedel-Kruse, Sartori:16}. Coordinated beating has been hypothesised considering different mechanisms such as dynein's activity regulation through local axonemal curvature \cite{Brokaw_71,Hines-Blum, Hines-Blum2, Sartori:16}, due to the presence of a transverse force (t-force) acting on the axoneme \cite{Lindemann} and by shear displacements \cite{Camalet,Camalet_NJP,Riedel-Kruse}. Other studies also examined the dynamics of flagellar beating by prescribing its internal activity \cite{Fu,Gadelha1} or by considering a self-organized mechanism independent of the specific molecular details underlying the collective action of dyneins \cite{Camalet,Camalet_NJP, Hilfinger}. In particular, the latter approach, although general from a physics perspective, it does not explicitly incorporate dynein kinetics along the flagellum, which has been shown to be crucial in order to understand experimental observations on sperm flagella \cite{Brokaw_1999,Brokaw_2014,Dynein_review}. Load-accelerated dissociation of dynein motors was proposed as a mechanism for axonemal sliding control, and was successfully used to infer the mechanical properties of motors from bull sperm flagella \cite{Riedel-Kruse}. In contrast, dynamic curvature regulation has been recently proposed to account for {\it Chlamydomonas} flagellar beating \cite{Sartori:16}. In the previous studies, linearized solutions of the models were fit to experimental data; however, it is unclear that such results still hold at the nonlinear level. Recent studies also investigated the emergence and saturation of unstable modes for different dynein control models \cite{Bayly2014,Bayly2015}; however, saturation of such unstable modes was not self-regulated, but achieved via the addition of a nonlinear elastic contribution in the flagellum constitutive relation. Nevertheless, predictions on how dynein activity influences the selection of the beating frequency, amplitude and shape of the flagellum remain elusive. Here,  we provide a microscopic bottom-up approach and consider the intrinsic nonlinearities arising from the coupling between dynein activity and flagellar shape, regarding the eukaryotic flagellum as a generalized Euler-elastica filament bundle \cite{Gadelha_PNAS}. This allows a close inspection on the onset of the flagellar bending wave instability, its transient dynamics and later saturation of unstable modes, which is solely driven by the nonlinear interplay between the flagellar shape and dynein kinetics.  \\

We first derive the governing nonlinear equations using a load-accelerated feedback mechanism for dynein along the flagellum. The linear stability analysis is presented, and eigenmode solutions are obtained similarly to Refs. \cite{Riedel-Kruse,Bayly2015}, to allow analytical progress and pedagogical understanding. The nonlinear dynamics far from the Hopf bifurcation is studied numerically and the resulting flagellar shapes are further analyzed using principal component analysis \cite{Werner,Jolliffe}. Finally, bending initiation and transient dynamics are studied subject to different initial conditions.  

\section{\label{sec:2} Continuum flagella equations}

\begin{figure}[t!]
\centering
\includegraphics[scale=0.70]{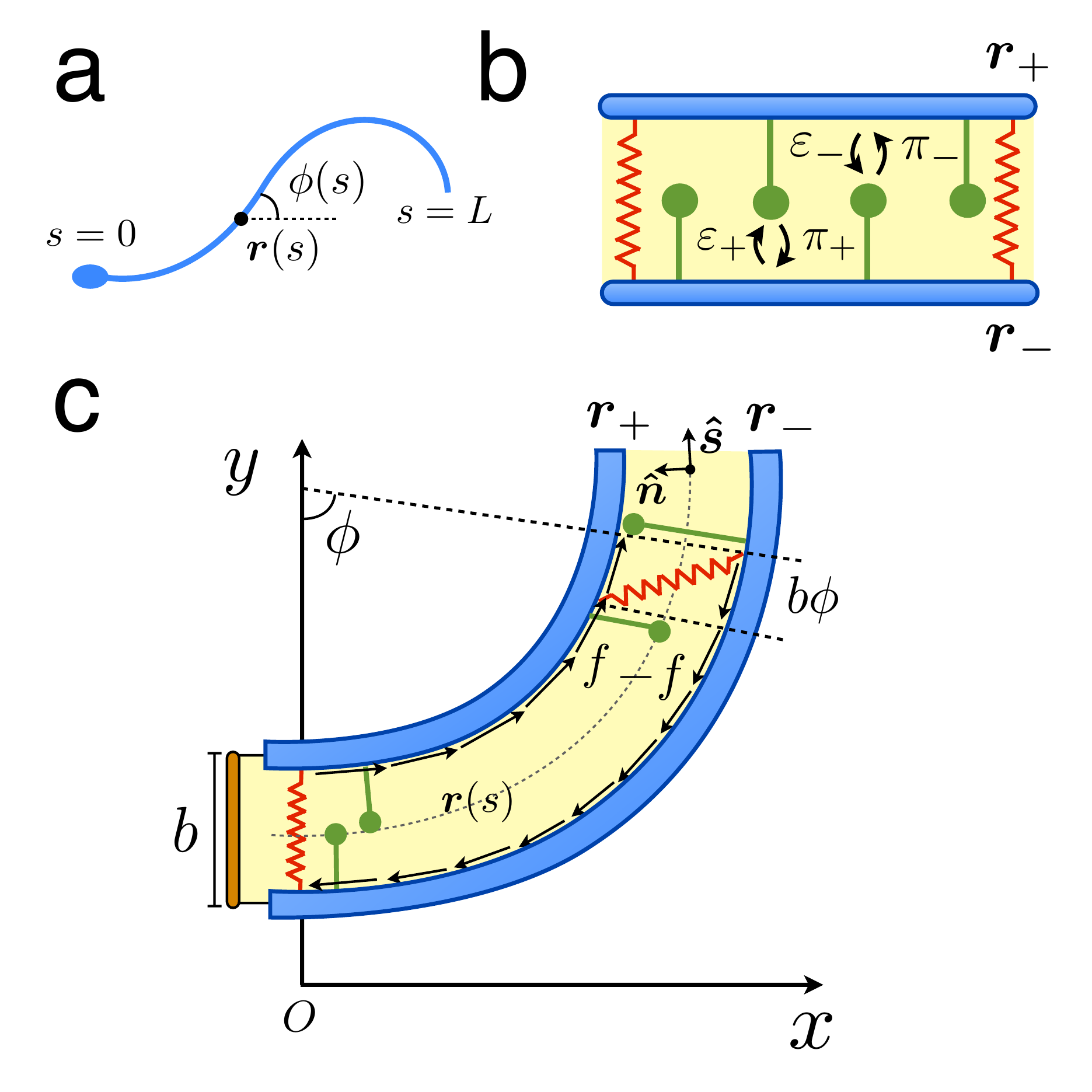}
\caption{(Online version in colour) Schematic view of the system a) Two-dimensional representation of a flagellum, where the centerline $\boldsymbol{r}(s)$ and tangent angle $\phi(s)$ of the flagellum are parametrized by the arc length parameter $s$. b) Passive (springs) and active (dynein motors) internal structures in the axoneme. Dyneins in the (+) and (-) filaments compete in a tug-of-war and bind/unbind from filaments with rates $\pi_{\pm}$ and $\varepsilon_{\pm}$ respectively. c) The flagellum as a two-filament bundle:  two polar filaments are separated by a small gap of size $b \ll L$. The presence of nexin crosslinkers and dynein motors generates a total force density $f(s)$ along the bundle. }
\label{fig1}
\end{figure}

We consider a filament bundle composed of two polar filaments subjected to planar deformations. Each filament is modelled as an inextensible, unshearable, homogeneous elastic rod, for which the bending moment is proportional to the curvature and the Young modulus is $E$. The filaments are of length $L$ and separated by a constant gap of size $b$, where $b \ll L$ (Fig. \ref{fig1}c). We define a material curve describing the shape of the filament bundle centerline as $\boldsymbol{r}(s,t)$. The positions of each polar filament forming the bundle read $\boldsymbol{r}_{\pm}=\boldsymbol{r} \pm (b/2) \boldsymbol{\hat{n}}$, with the orientation of the cross-section at distance $s$ along its length defined by the normal vector to the centerline $\boldsymbol{\hat{n}}=-\sin \phi \, \boldsymbol{\hat{\imath}}+\cos \phi \, \boldsymbol{\hat{\jmath}}$, being $\phi \equiv \phi(s,t)$ the angle between the tangent vector $\boldsymbol{\hat{s}} \equiv  \partial_s \boldsymbol{r} \equiv \boldsymbol{r}_s$ and the $\boldsymbol{\hat{\imath}}$ direction (taken along the {\it x} axis). The subscripts (+) and (-) refer to the upper and lower filaments, respectively (Fig. \ref{fig1}c). The shape of the bundle is given at any time by the expression:
\begin{equation}
\boldsymbol{r}(s,t)=\boldsymbol{r}(0,t)+\int_{0}^{s} ( \cos \phi, \sin \phi) ds'
\label{Eq1}
\end{equation}
The geometrical constraint of the filament bundle, induces an arc length mismatch $\Delta(s,t)$, denoted as sliding displacement:
\begin{equation}
\Delta(s,t) = \int_0^{s} (| \partial_s \boldsymbol{r}_{-}|- | \partial_s \boldsymbol{r}_{+}|) ds' = b(\phi-\phi_0)
\label{Eq2}
\end{equation}
where $\phi_0 \equiv \phi(0,t)$. For simplicity, we have set any arc length incongruity between the two filaments at the base to zero and we will consider the filaments clamped at the base. A similar approach can be used to include basal compliance and other types of boundary conditions at the base (e.g. pivoting or free swimming head). Here we centre our study on the nonlinear action of motors along the flagellum. We aim to study the active and passive forces generated at each point along the arc length of the filament bundle. We define $\boldsymbol{f}(s,t)=f(s,t) \boldsymbol{\hat{s}}$ as the total internal force density generated at $s$ at time $t$ on the plus-filament due to the action of active and passive forces (see Fig. \ref{fig1}). By virtue of the action-reaction law, the minus-filament will experience a force density $-\boldsymbol{f}$ at the same point. Next, consider that $N$ dyneins are anchored at each polar filament in a region $l_c$ around $s$, where $l_c$ is much smaller than the length of the flagellum $L$ and much larger than the length of the regular intervals dyneins are attached to along the microtubule doublets. We shall call $l_c$ the `tug-of-war' length. We define $n_{\pm}(s,t)$ as the number of bound dyneins in a region of size $l_c$ around $s$ at time $t$ which are anchored in the plus- or minus-filament respectively. 
We consider a tug-of-war at each point $s$ along the flagellum with two antagonistic groups of $N$ dyneins. The elastic sliding resistance between the two polar filaments exerted by nexin crosslinkers is assumed to be Hookean with an elastic modulus $K$. Thus, the internal force density $f(s,t)$ reads:
\begin{equation}
f(s,t)=\rho (n_{+}F_{+}+ n_{-}F_{-})-K \Delta 
\label{Eq3}
\end{equation}
where $\rho \equiv l_c^{-1}$ is the density of tug-of-war units along the flagellum and $F_{\pm}(s,t)$ is the load per motor each group of dyneins experiences due to the action of the antagonistic group. The stresses on the filament bundle are given by a resultant contact force $\boldsymbol{N}(s,t)$ and resultant contact moment $\boldsymbol{M}(s,t)$ acting at the point $\boldsymbol{r}(s,t)$. The internal force density $f(s,t)$ only contributes to the internal moment of the bundle $\boldsymbol{M}(s,t)=M \boldsymbol{\hat{k}}$, {\bl where $\boldsymbol{\hat{k}} = \boldsymbol{\hat{\jmath}} \times \boldsymbol{\hat{\imath}}$}, such that $M(s,t)$ reads:
\begin{equation}
M(s,t)=E_b \phi_s -bF
\label{Eq4}
\end{equation}
where $F(s,t)=\int_{s}^{L} f(s',t)ds'$, provided that $\partial_s \phi_{ \pm} \approx \partial_s \phi$ for bundles characterized by $b \ll L$. The combined bending stiffness of the filament bundle is given by $E_b = 2EI$, where $I$ is the second moment of the area of the external rods. \\ 

Dynein kinetics is modelled by using a minimal two-state mechanochemical model with states $k=1,2$, corresponding to microtubule bound or unbound dyneins, respectively.  Since the sum of bound and unbound motors at $s$ remains constant at all times, we only study the plus and minus bound motor distributions $n_{\pm}(s,t)$. Dyneins bind with rates $\pi_{\pm}$ and unbind with rates $\varepsilon_{\pm}$ (Fig. \ref{fig1}b). The corresponding bound motor population dynamics reads:
\begin{equation}
\partial_t n_{\pm}= \pi_{\pm}-\varepsilon_{\pm}
\label{Eq5}
\end{equation}
The binding/unbinding rates are given by $\pi_{\pm}=\pi_0(N-n_{\pm})$, $\varepsilon_{\pm}= \varepsilon_0 n_{\pm} \exp(\pm F_{\pm}/f_c)$ where $\varepsilon_0$ and $\pi_0$ are constant rates and $f_c$ is the characteristic unbinding force. {\bl Motivated by previous studies on the collective action of molecular motors \cite{Riedel-Kruse, Muller}, we assume an exponential dependence of the unbinding force on the resulting load}. By considering that dyneins fulfill a linear velocity-force relationship with stall force $f_0$ and velocity at zero load $v_0$, the loads are given by $F_{\pm}(s,t)=\pm f_0 ( 1\mp \Delta_t/v_0 )$. {\bl The stall force $f_0$ is defined as the absolute value of the load a motor experience at stall ($\Delta_t=0$)}. Substituting the different definitions, the internal force density $f(s,t)$ reads:
\begin{equation}
\bl{f(s,t)= f_0 \rho \left( \bar{n}-\frac{n' \Delta_t}{v_0}  \right)-K \Delta}
\label{Eq6}
\end{equation}
where $\bar{n} \equiv n_{+}-n_{-}$ {\bl and $n' \equiv n_{+}+n_{-}$}. For simplicity, we will derive the equations governing the tangent angle $\phi$ in the limit of small curvature (but possibly large amplitudes) such that tangential forces can be neglected. The derivation for arbitrary large curvature is also presented in the electronic Supplementary Text. Using resistive force theory in the limit of small curvature, we only consider normal forces along the flagellum obtaining $\zeta_{\perp} \phi_t = - M_{sss}$, where $\zeta_{\perp}$ is the normal drag coefficient \cite{Machin, Gray_Hancock}. Combining the last expression with Eq. \ref{Eq4} we have:
\begin{equation}
\zeta_{\perp} \phi_t = -E_b \phi_{ssss} - bf_{ss} 
\label{Eq7}
\end{equation}
Hereinafter we switch to dimensionless quantities while keeping the same notation. We non-dimensionalize the arc length with respect to the length scale $L$, time with respect to the correlation time of the system $\tau_0=1/(\varepsilon_0+\pi_0)$, motor number with respect to $N$, internal force density with respect to $f_0 \rho N$ and sliding displacement with respect to $b$. The correlation time defines how fast the motors will respond to a change in load. We also define $\mbox{Sp} = L (\zeta_{\perp}/E_b\tau_0)^{1/4}$, $\mu \equiv Kb^2L^2/E_b$, $\mu_a=b f_0 \rho N L^2/E_b$ and $\zeta \equiv b/v_0 \tau_0$. The sperm number $\mbox{Sp}$ characterizes the relative importance of bending forces to viscous drag. The parameter $\mu$ measures the relative importance of the sliding resistance compared with the bending stiffness \cite{Gadelha_PNAS}. On the other hand, the parameter $\mu_a$ denotes the activity of dyneins, measuring the relative importance of motor force generation compared with the bending stiffness of the bundle. Finally $\zeta$ denotes the ratio of the bundle diameter and the characteristic shear induced by the motors. The dimensionless sperm equation in the limit of small curvature reads \cite{Lauga, Camalet, Gadelha1}:
\begin{equation}
\mbox{Sp}^4 \phi_t = - \phi_{ssss} - \mu_a f_{ss}
\label{Eq8}
\end{equation}
where in our case the dimensionless internal force density $f(s,t)$ takes the form:
\begin{equation}
\bl{f(s,t)=\bar{n}-\zeta n' \Delta_t -\frac{\mu}{\mu_a}\Delta}
\label{Eq9}
\end{equation}
and $\Delta= \phi-\phi_0$. Since the flagellar base is clamped, without loss of generality, we set $\phi_0=0$. Combining Eqs. \ref{Eq8} and \ref{Eq9} we obtain the nonlinear dynamics for the tangent angle:
\begin{equation}
\bl{\mbox{Sp}^{4} \phi_t = - \phi_{ssss} + \mu \phi_{ss} -\mu_a \bar{n}_{ss}+\mu_a \zeta [n_{ss}'\phi_t + 2 n_s' \phi_{ts}+n'\phi_{tss}]}
\label{Eq9_2}
\end{equation}
In the absence of dynein activity, the last expression reduces to the dynamics of an elastic filament bundle with sliding resistance forces \cite{Gadelha_PNAS}. Notice that this expression is obtained considering the sliding mechanism and a linear velocity-force relationship for dyneins, but it is independent of dynein kinetics. On the other hand, the dimensionless form of the bound motor population dynamics $n_{\pm}$ reads:
\begin{equation}
\partial_t n_{\pm}=\eta(1-n_{\pm})-(1-\eta) n_{\pm} \exp [\bar{f}(1\mp \zeta \phi_t)]
\label{Eq10}
\end{equation}
where $\eta \equiv \pi_0 /(\pi_0 +\varepsilon_0)$ is the duty ratio of the motors and $\bar{f} \equiv f_0/f_c$ dictates the sensitivity of the unbinding rate on the load. In the Supplementary Text we include a table with a summary of all the variables and parameters in the model with their corresponding symbols. 

\section{\bl{\label{sec:2b}  Parameter values}}

{\bl We present the choice of parameters based on experimental studies on sperm flagella and on the green algae {\it Chlamydomonas}. We first discuss the passive properties of a flagellum. The typical length of a human flagellum is $L \simeq 50$ $\mu$m and the axonemal diameter is found to be $b \simeq 200$ nm \cite{Gaffney_review}. The bending stiffness of the filament bundle has been reported to be $E_b \simeq  0.9 \cdot 10^{-21}$ N$\cdot$m$^2$ for sea-urchin sperm \cite{Gadelha_PNAS,Gaffney_review} and $E_b \simeq  1.7 \cdot 10^{-21}$ N$\cdot$m$^2$ for bull sperm \cite{Riedel-Kruse}. On the other hand, the interdoublet elastic resistance from demembranated flagellar axonemes of {\it Chlamydomonas} yields an estimated spring constant $2 \cdot 10^{-3}$ N/m for 1 $\mu$m of axoneme \cite{Minoura}, thus $K \simeq 2 \cdot 10^{3}$ N/m$^{2}$. Finally, typical medium viscosities for sperm flagella range from $\zeta_{\perp} \simeq 10^{-3}$ Pa$\cdot$s in low viscous media to $\zeta_{\perp} \simeq 1$ Pa$\cdot$s in high viscous media \cite{Gaffney_review,Gadelha1}. \\
 
Next, we discuss the mechanochemical parameters associated to axonemal dynein. Axonemal dynein are subdivided in inner and outer arms depending on its position in the axoneme, and can be found in heterodimeric and monomeric forms \cite{Dynein_review}. For the sake of simplicity, we consider identical force generating dynein motor domains acting along the flagellum. The total number of motor domains in a beating flagellum has been estimated to be $\simeq 10^{5}$ \cite{Ma,Nicastro}. The stall force has been found in the range $f_0 \simeq 1-5$ pN \cite{Sakakibara,Hirakawa}. Following Ref. \cite{Riedel-Kruse} we choose the characteristic unbinding force for dynein such that $\bar{f}=2$. Axonemal dynein is characterized by a low duty ratio estimated to be $\eta \simeq 0.14$ and speeds at zero load in the range $v_0 \simeq 5-7 \, \mu \mbox{m/s}$ \cite{Sakakibara, Howard}. The characteristic time scale of dynein kinetics sets the order of magnitude of the beating frequency in our model. We choose an estimated value of $\tau_0 \simeq 50$ ms \cite{Bayly2015,Howard}, which corresponds to a frequency of $\simeq 10$ Hz, comparable to the case of human sperm \cite{Gaffney_review}. Finally, we need to estimate $\rho$ and $N$. Considering the length of the human sperm flagellum ($L \simeq 50$ $\mu$m) we obtain $\simeq 2 \cdot 10^{3}$ motors/$\mu \mbox{m}$. In our description, we divide the axoneme in two regions with corresponding dynein teams. Therefore, we have  $\rho N \simeq 10^{3}$ motors/ $\mu \mbox{m}$. In order to find $\rho$, we need to choose a criterion to decide the typical length scale or `tug-of-war' length $l_c = \rho^{-1}$ in our coarse-grained description. The typical length scale in the system is given by $l_c \sim L/\sqrt{\mu_a}=\sqrt{E_b/bf_0 \rho N}$ (see Linear stability analysis). Using the previous parameters we get $l_c \sim 1$ $\mu$m and therefore $\rho \sim 1$ $\mu$m$^{-1}$ and $N \sim 10^{3}$. Hence, we obtain $\mbox{Sp} \simeq 5-20$, $\mu \simeq 50-100$, $\mu_a \sim 10^{3}$ and $\zeta \sim 1$. The motor activity is studied in a broad range $\mu_a \simeq (2-6) \cdot 10^{3}$ since it plays the role of the main control parameter in our study. }

\section{\label{sec:3} Linear stability analysis}

{\bl In this section, we perform a linear stability analysis of Eqs. \ref{Eq9_2} and \ref{Eq10}}. The nonmoving state is characterized by $\phi = 0$ and $n_{\pm} = n_0 \equiv \pi_0/(\pi_0+\varepsilon_0 e^{\bar{f}})$. This means that the flagellum is aligned with respect to the $x$-axis and the number of plus and minus bound motors is constant in space and time. 
For the linear stability analysis, we consider the perturbed variables around the base state as $\phi= \delta \phi$ and $n_{\pm}=n_0 + \delta n_{\pm}$. Introducing the modulation $\delta n \equiv \delta n_{+}=-\delta n_{-}$ around $n_0$ and considering $\bar{f} \zeta \phi_t \ll 1$ we obtain:
\begin{eqnarray}
\delta n_t &=& - \bar{\tau}^{-1} \delta n + (1-\eta) \zeta \bar{f} e^{\bar{f}}n_0 \phi_t  \label{Eq11A} \\
\mbox{Sp}^4 \phi_t &=& -\phi_{ssss} + \mu \phi_{ss} +2 \mu_a[\zeta n_0 \phi_{tss}-\delta n_{ss}] \label{Eq11B}
\end{eqnarray}
where $\bar{\tau} \equiv n_0/\eta$. We use the ansatzs $\phi=\tilde{\phi}(s)e^{\sigma t}+\mbox{c.c}$ and $\delta n= \delta \tilde{n}(s) e^{\sigma t}+\mbox{c.c}$, where $\sigma$ is a complex eigenvalue and $\mbox{c.c}$ accounts for complex conjugate. From Eq. \ref{Eq11A} we get $\delta \tilde{n}= \chi'(\sigma) \tilde{\phi}$, where $\chi'(\sigma)$ is a complex response function:
\begin{equation}
\chi'(\sigma) = \zeta \bar{f} n_0(1-n_0) \frac{ \sigma}{1+\sigma \bar{\tau}} 
\label{Eq12}
\end{equation}
Using Eq. \ref{Eq9} and considering $f=\tilde{f}(s)e^{\sigma t}+\mbox{c.c}$, we obtain $\tilde{f}=\chi(\sigma) \tilde{\phi}$, where $\chi(\sigma)$ is a second complex response function:
\begin{equation}
\chi(\sigma)=2 \zeta n_0\left[ \bar{f}(1-n_0)\frac{\sigma-\sigma^2 \bar{\tau}}{1-(\sigma \bar{\tau})^2}-\sigma \right] - \frac{\mu}{\mu_a}
\label{Eq13}
\end{equation}
The latter response functions generalize the work in Ref. \cite{Riedel-Kruse} for a complex eigenvalue $\sigma$ and are equivalent to results presented in Ref. \cite{Bayly2015}. With the ansatz $\tilde{\phi} \sim \delta \tilde{n} \sim e^{iqs}$ in Eq. \ref{Eq11B}, we obtain the characteristic equation $q^4-\bar{\chi} q^2 + \bar{\sigma} =0$, where $\bar{\chi} \equiv \mu_a \chi$, $\bar{\sigma} \equiv \sigma \mbox{Sp}^4$ {\bl and} $ \bar{\chi}, \bar{\sigma} \in \mathbb{C}$. Solving the characteristic equation we obtain four possible roots $q_i$,  $i=1, \ldots, 4$ and the eigenfunctions read:
 \begin{equation}
 \tilde{\phi}(s)=\sum_{j=1}^{4} \tilde{\Phi}_j e^{iq_js} 
 \label{Eq14}
 \end{equation}
where $q_j, \tilde{\Phi}_j \in \mathbb{C}$. Once $\tilde{\phi}$ is known, $\delta \tilde{n}(s) = \delta N \tilde{\phi}(s) \exp(i \Delta \theta)$ where $ \delta N=|\chi' |$ and $\Delta \theta=\arg(\chi')$. Therefore, the evolution of $\delta n$ is the same as for $\phi$ except for a phase shift $\Delta \theta$ and an overall change on the amplitude $\delta N$, which depends on $\chi'(\sigma)$. This result indicates the presence of a time delay between the action of motors and the response of the flagellum. Time delays commonly arise in systems where molecular motors work collectively \cite{Oriola_EPL}. Indeed, the regulation of active forces by the time delay of the curvature was proposed as a mechanism to generate travelling bending waves \cite{Brokaw_71}. 
\begin{figure}[h!]
\centering
\includegraphics[scale=0.60]{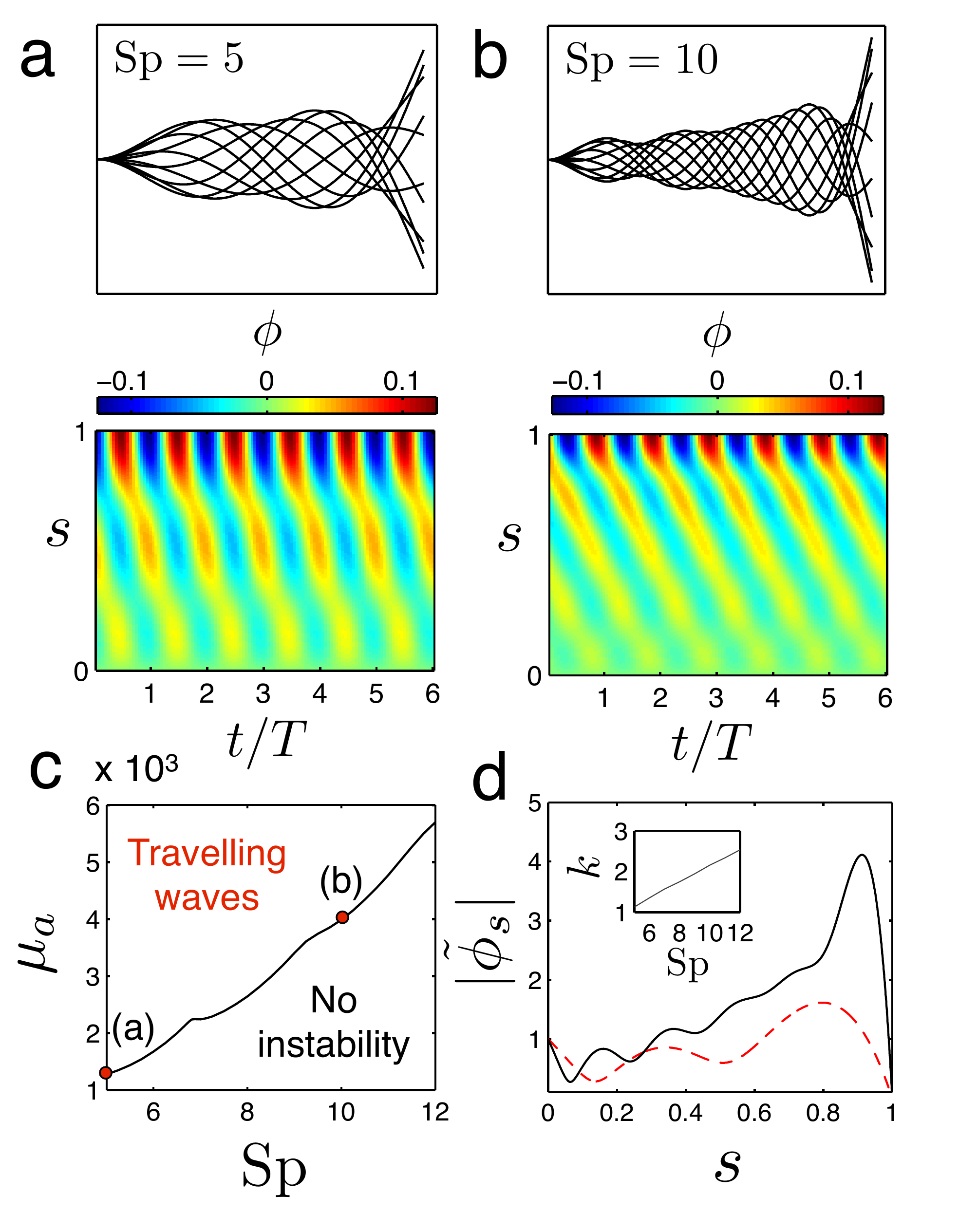}
\caption{(Online version in colour) Linear stability analysis. a,b) (Upper panels) Clamped head profile solutions corresponding to the marginal stability case (i.e. $\lambda_1=0$) for $\mbox{Sp}=5,10$. The beating cycles are divided in 10 frames. (Lower panels) Tangent angle kymographs where $T=2\pi/\omega_c$ is the period. Amplitudes and angles are shown in arbitrary units. c) Marginal stability curve for the clamped condition. Points (a) and (b) in parameter space correspond to the profiles in a) and b). d) Curvature modulations as a function of the arc length for $\mbox{Sp}=5$ (dashed line) and $\mbox{Sp}=12$ (solid line). Inset: Wavenumber defined as $k \equiv  \max |{q_i}|/2\pi$  as a function of $\mbox{Sp}$. $\mu=50$, $\zeta=0.4$, $\eta=0.14$, $\bar{f}=2$.  }
\label{fig2}
\end{figure}
In order to find $\tilde{\phi}(s)$, we need to impose the four boundary conditions, obtaining a linear system of equations for $\tilde{\Phi}_j$, $j=1,\ldots,4$. {\bl The procedure to obtain the set of boundary conditions for a clamped head is detailed in the Supplementary Text. The boundary conditions read $\tilde{\phi}(0)=0$, $\tilde{\phi}_{sss}(0)=-\bar{\chi} \tilde{\phi}_s(0)$, $\tilde{\phi}_s(1)=0$ and $\tilde{\phi}_{ss}(1)=-\bar{\chi} \tilde{\phi}(1)$}. By setting the determinant of the system to zero, we find the set of complex eigenvalues $\sigma_n$, with the corresponding growth rates $\lambda_n=\mbox{Re}[\sigma_n]$ and frequencies $\omega_n= \mbox{Im}[\sigma_n]$, which satisfy the boundary conditions, where $\lambda_n, \omega_n \in \mathbb{R}$. We order the set of different eigenvalues according to its growth rate $\lambda_{n+1}< \lambda_n$, such that the first one has the largest growth rate $\lambda_1$. Defining $\boldsymbol{u}=( \phi, \delta n)^{T}$, the general solution of the system reads:
\begin{eqnarray}
\boldsymbol{u}(s,t) &=&\sum_n A_n 
\begin{pmatrix} \tilde{\phi}_n\\ \delta \tilde{n}_n \end{pmatrix}
e^{\lambda_n t} e^{i \omega_nt} + \mbox{c.c.} 
 \label{Eq15}
\end{eqnarray}
where $A_n \in \mathbb{C}$ are free amplitude parameters. For $\lambda_n < 0$, $\forall n$, solutions decay exponentially to the nonmoving state. On the other hand, when $\lambda_1$ becomes positive, {\bl in the range of parameters studied}, the system undergoes a Hopf bifurcation and solutions follow an exponential growth, oscillating with frequency $\omega_1$. Next, we study the marginal stable solutions, i.e. when the maximum growth rate equals zero ($\lambda_1=0$). For this, we define the critical frequency of oscillation as $\omega_c \equiv |\omega_1|$. Travelling waves propagate from tip to base, a feature already reported for the clamped type boundary condition \cite{Camalet,Camalet_NJP,Bayly2015}. In Fig. \ref{fig2}c, the marginal stability curve in phase space is shown. Intuitively, as $\mbox{Sp}$ is increased the travelling instability occurs for higher motor activity $\mu_a$ and the critical frequency of oscillation $\omega_c$ follows a non-monotonic decrease (see Supplementary Figure S1). For low viscosity ($\mbox{Sp}=5$) the wave propagation velocity is slightly oscillatory whereas for high viscosity ($\mbox{Sp}=10$) it becomes more uniform (Fig. \ref{fig2}a and b, lower panels). These results are in agreement with studies on migrating human sperm, where in the limit of high viscosity waves propagated approximately at constant speed \cite{Smith}. For high viscosity, curvature tends to increase from base to tip, finally dropping to zero due to the zero curvature boundary condition at the tail (see Fig. \ref{fig2}d and Supplementary Text). This modulation is consistent with experimental studies on human sperm, which show viscosity modulation of the bending amplitude \cite{Smith}.
In the latter study; however, the effect is more pronounced possibly due to external elastic reinforcing structures found along the flagellum of mammalian species, as well as other nonlinear viscoelastic effects. Defining $k \equiv  \max |{q_i}|/2\pi$ as the characteristic wavenumber, we obtain that it increases almost linearly with \mbox{Sp} (Fig. \ref{fig2}d, inset). Similar results can be obtained using other definitions for $k$, for example using the covariance matrix (see Principal component analysis section). \\
\section{\label{sec:4} Nonlinear flagellar dynamics} 
\begin{figure}[h!]
\centering
\includegraphics[scale=0.65]{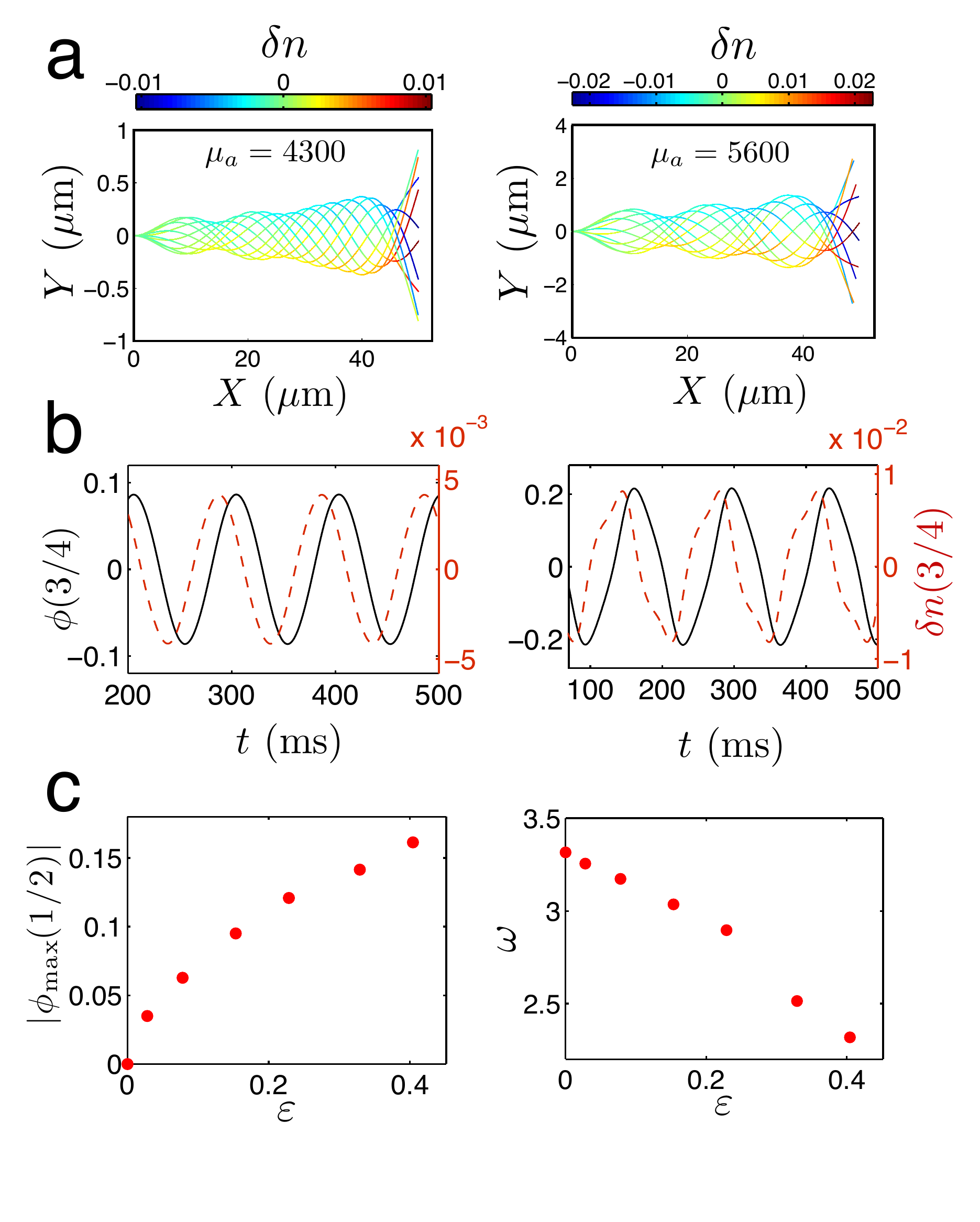}
\caption{(Online version in colour) Nonlinear analysis. a) Nonlinear flagella steady state profiles for $\mu_a=4300$ (left)  and $\mu_a=5600$ (right), considering the respective eigenmodes as initial conditions. The beating cycles are divided in 10 frames as in Fig. \ref{fig2}a,b (upper panels). b) $\phi$ (solid lines) and $\delta n$ (dashed lines) evaluated at $s=3/4$ for the profiles in (a) respectively.  c) Maximum absolute tangent angle evaluated at $s=1/2$ and dimensionless frequency $\omega$ as a function of the relative distance to the bifurcation $\varepsilon = (\mu_a-\mu_a^{c})/\mu_a^{c}$. $\mbox{Sp}=10$, $\mu=50$, $\eta=0.14$, $\zeta=0.4$, $\bar{f}=2$, $L=50$ $\mu$m and $\tau_0=50$ ms. }
\label{fig3}
\end{figure}

In this section, we study the nonlinear dynamics of the flagellum in the limit of small curvature by numerically solving Eqs. \ref{Eq9_2} and \ref{Eq10} using a second-order accurate implicit-explicit numerical scheme (see Supplementary Text). The unstable modes presented in Section \ref{sec:3} follow an initial exponential growth and eventually saturate at the steady state due to the nonlinearities in the system. In Fig. \ref{fig3} two different saturated amplitude solutions are shown. Fig. \ref{fig3}a (left) corresponds to a case where the system is found close to the Hopf bifurcation, whereas Fig. \ref{fig3}a (right) corresponds to a regime far from the bifurcation. We notice that the marginal solution obtained in the linear stability analysis (Fig. \ref{fig2}b, upper panel) gives a very good estimate of the nonlinear profile close to the bifurcation point, although it does not provide the magnitude of $\phi$ or $\delta n$. Frequencies are $\sim 10$ Hz and maximum amplitudes are found to be small, around $\simeq 4 \%$ of the total flagellum length. However, the oscillation amplitude for high motor activity is more than double in respect to the case of low activity (Fig. \ref{fig3}a). The colour code in Fig. \ref{fig3}a indicates the value of the semi-difference of plus- and minus-bound motors $\delta n$. Plus-bound motors are predominant in regions of positive curvature ($\phi_s>0$) along the flagellum and vice-versa.
 Despite the low duty ratio of dynein motors \cite{Howard}, $\sim 2 \%$ bound dyneins along the flagellum are sufficient to produce micrometer-sized amplitude oscillations. The full flagella dynamics corresponding to Fig. \ref{fig3}a are provided in the Supplementary Movies 1 and 2. \\
 
In Fig. \ref{fig3}b, the time evolution of $\phi$ and $\delta n$ is shown at $s=3/4$ for the cases in Fig. \ref{fig3}a, respectively. As mentioned in Section \ref{sec:3}, the tangent angle $\phi$ is delayed respect to $\delta n$, and the time delay is not considerably affected by motor activity. Close to the instability threshold, both signals are very similar since the system is found near the linear regime; however, far from threshold, both signals greatly differ. For high motor activity, both the tangent angle and the fraction of bound dyneins at certain points along the flagellum exhibit cusp-like oscillations (Fig. \ref{fig3}b, right). This behaviour is typical of molecular motor assemblies working far from the instability threshold \cite{Julicher_spontaneous}.  Despite the signals $\mbox{S}$ in Fig. \ref{fig3}b (right) are nonlinear, they conserve the symmetry $\mbox{S}(t+T/2) = -\mbox{S}(t)$, {\bl where $T$ is} the period of the signal. This is a consequence of both plus and minus motor populations being identical, a property also found in spontaneous oscillations of motor assemblies \cite{Julicher_spontaneous, Oriola_EPL}. Finally, in Fig. \ref{fig3}c we study how the amplitude and frequency of the oscillations vary with the relative distance from the bifurcation point $\varepsilon \equiv (\mu_a-\mu_a^{c})/\mu_a^{c}$. For small $\varepsilon$, the maximum absolute value of the tangent angle seems to follow a square root dependence, characteristic of supercritical Hopf bifurcation; however, in the strongly nonlinear regime the curve deviates from this trend. On the other hand, the beating frequency decreases for increasing activity. {\bl The only possibility to increase $\mu_a$ keeping the other dimensionless parameters fixed is by increasing $\rho N$; hence, the larger each dynein team becomes, the larger the activity.}  Consequently, the necessary time to unbind a sufficient number of dynein motors to drive the instability increases, leading to a lower beating frequency.  

\subsection{\label{sec:5} Principal component analysis} 

\begin{figure}[h!]
\centering
\includegraphics[scale=0.55]{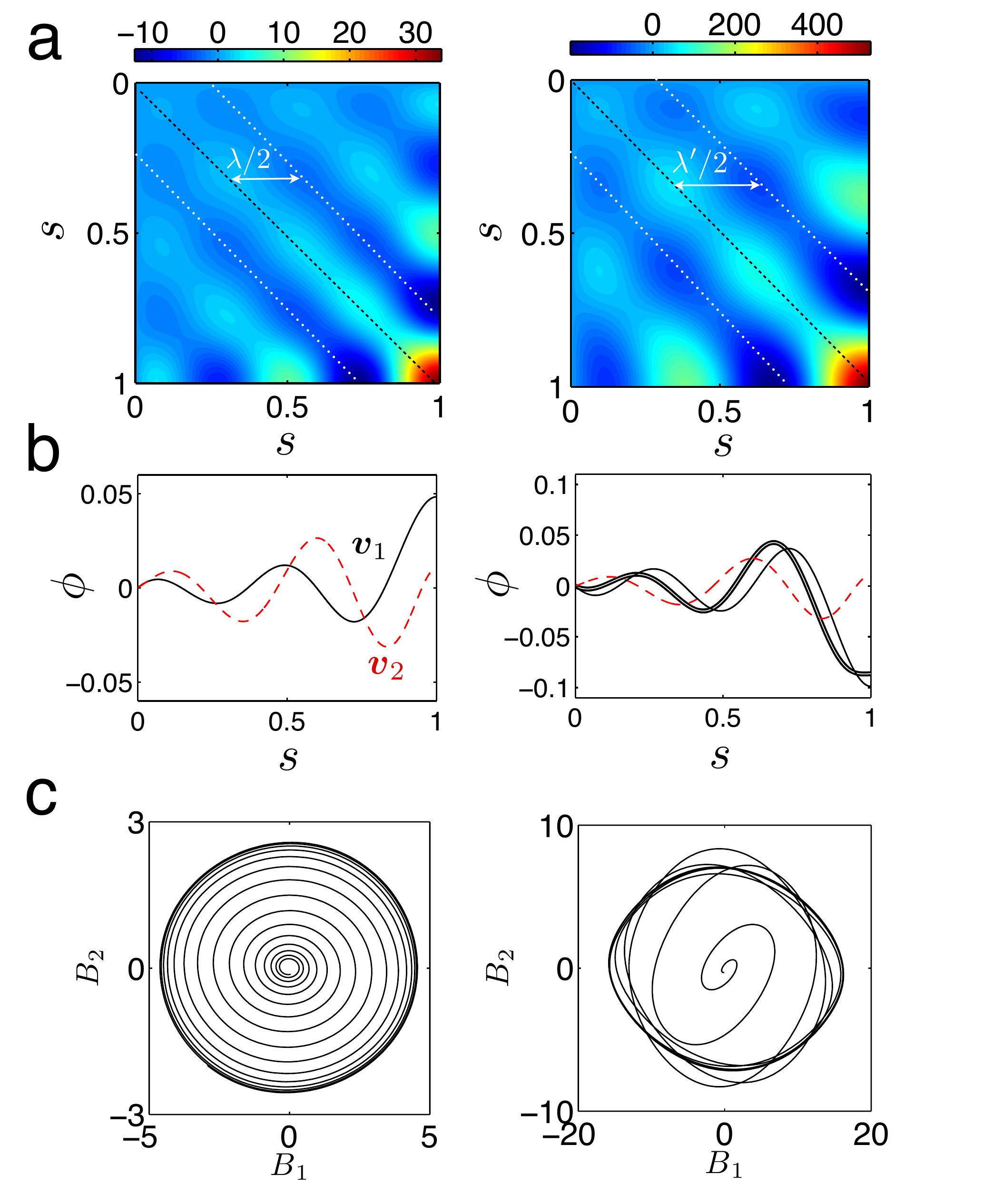}
\caption{(Online version in colour) Principal component analysis of flagellar beating. a) Covariance matrix for $\mu_a=4300$ (left) and $\mu_a=5600$ (right). We can identify characteristic wavelengths $\lambda, \lambda'$ from negative long-range correlations in $\mbox{C}$. Notice $\lambda'> \lambda$ and the number of local maxima decreases when $\mu_a$ is increased. b) (Left) Two principal shape modes $v_1,v_2$ (solid and dashed lines, respectively), corresponding to the two maximum eigenvalues of the covariance matrix in Fig \ref{fig4}a (left). (Right) The flagellar shape at time $t=733$ ms (thick solid line) is reconstructed (white line) by a superposition of the two principal shape modes $v_1,v_2$ in Fig. \ref{fig4}b (left) fitting the scores $B_1,B_2$. c) Flagellar dynamics in a reduced two-dimensional shape space for $\mu_a=4300$ (left) and $\mu_a=5600$ (right). Elliptic limit cycles are rescaled to better appreciate the distortion due to the nonlinear terms. $\mbox{Sp}=10$, $\mu=50$, $\eta=0.14$, $\zeta=0.4$. }
\label{fig4}
\end{figure}

In this section, we study the obtained nonlinear solutions using principal component analysis \cite{Werner,Jolliffe}. This technique treats the flagellar shapes as multi-feature data sets, which can be projected to a lower dimensional space characterized by principal shape modes. Here we will analyze the numerically resolved data following Ref. \cite{Werner} to study sperm flagella. We discretize our flagella data with time-points $t_i$, $i=1, \ldots, p$ and $M=4 \cdot 10^3$ intervals corresponding to points $s_j=(j-1)/M$, $j=1, \ldots, r$ along the flagellum, with $r=M+1$. We construct a measurement matrix $\Phi$ of size $p \times r$ for the tangent angle where $\Phi_{ij}=\phi(s_j,t_i)$. This matrix represents a kymograph of the flagellar beat.  We define the $r \times r$ covariance matrix as $\mbox{C}= (\Phi-\bar{\Phi})^{T} (\Phi-\bar{\Phi})$, where $\bar{\Phi}=[\boldsymbol{\bar{\phi}}; \ldots; \boldsymbol{\bar{\phi}}]$ with all rows equal to $\boldsymbol{\bar{\phi}}=[\bar{\phi}(s_1), \ldots, \bar{\phi}(s_r)]$, {\bl where $\bar{\phi}(s_i)$ is} the mean tangent angle at $s_i$. The covariance matrix $\mbox{C}$ is shown for $\mu_a=4300$ (Fig. \ref{fig4}a, left) and $\mu_a=5600$ (Fig. \ref{fig4}a, right). In Fig. \ref{fig4}a (left), we find negative correlation between tangent angles that are a distance $\lambda/2$ apart. Hence, a characteristic wavelength $\lambda$ can be identified in the system, which manifests as a long-range correlation in the matrix $\mbox{C}$. On the other hand, strong positive correlations around the main diagonals correspond to short-ranged correlations mainly due to the bending stiffness of the bundle \cite{Werner}. The number of local maxima along the diagonals in $\mbox{C}$ decreases from $\mu_a=4300$ to $\mu_a=5600$, and at the same time $\lambda'> \lambda$. Hence, an increase in motor activity slightly increases the characteristic wavelength while decreasing the number of local maxima, which is related to the characteristic wavenumber. Employing an eigenvalue decomposition of the covariance matrix, we can obtain the eigenvectors $\boldsymbol{v}_1, \ldots, \boldsymbol{v}_r$ and their corresponding eigenvalues $d_1, \ldots, d_r$, such that $\mbox{C} \boldsymbol{v}_j= d_j \boldsymbol{v}_j$. Without loss of generality, we can sort the eigenvalues in descending order $d_1 \geq \ldots \geq d_r$. We find that the first two eigenvalues capture $>99 \%$ variance of the data. This fact indicates that our flagellar waves can be suitably described in a two-dimensional shape space, since they can be regarded as single-frequency oscillators. Each flagellar shape $\Phi_i=[\phi(s_1,t_i), \ldots, \phi(s_r,t_i)]$ can be expressed now as a linear combination of the eigenvectors $\boldsymbol{v}_{k}$ \cite{Werner}:
\begin{equation}
\Phi_i=\boldsymbol{\bar{\phi}}+\sum_{k=1}^{r} B_{k}(t_i)\boldsymbol{v}_k
 \label{Eq16}
\end{equation}
where $B_k$ are the shape scores computed by a linear least-square fit. In Fig. \ref{fig4}b (left), the two first eigenvectors $v_1,v_2$ are shown for $\mu_a=4300$. In Fig. \ref{fig4}b (right), the flagellar shape at a certain time (thick solid line) is reconstructed (white line) by using a superposition of the two principal shape modes $v_1,v_2$ (solid and dashed lines, respectively) and fitting the scores $B_1, B_2$. Finally, in Fig. \ref{fig4}c we show the shape space trajectories beginning with small amplitude eigenmode solutions. While close to the bifurcation the limit cycle is elliptic (Fig. \ref{fig4}c, left), far from the bifurcation the limit cycle becomes distorted (Fig. \ref{fig4}c, right). Elliptic limit cycles were also found experimentally for bull sperm flagella \cite{Werner}. Hence, as found in Section \ref{sec:4}, motor activity in the nonlinear regime significantly affects the shape of the flagellum when compared with the linear solutions, which only provide good estimates sufficiently close to the Hopf bifurcation. 

\subsection{\label{sec:6} Bending initiation and transient dynamics} 
\begin{figure}[h!]
\centering
\includegraphics[scale=0.55]{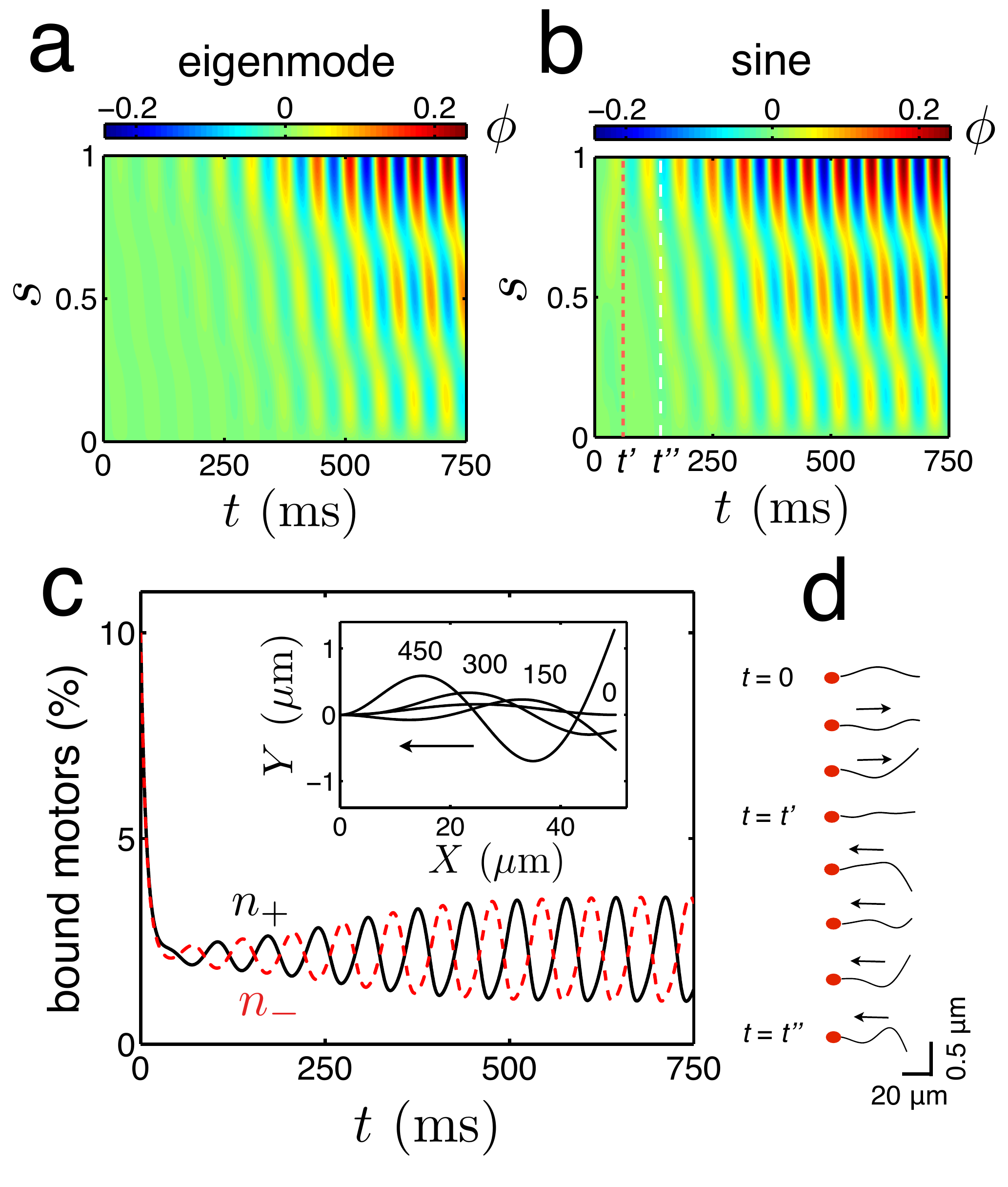}
\caption{(Online version in colour) Bending initiation and transient dynamics. a) Tangent angle kymograph for an eigenmode initial condition b) Tangent angle kymograph for a sinusoidal initial condition for the tangent angle with $n_{+}(0)=n_{-}(0)=0.1$. c) Bound motor time evolution for the plus (solid line) and minus (dashed line) dynein populations at $s=9/10$ for the case of a sinusoidal initial condition. Inset: Flagella profiles at different times in ms. d) Snapshots of the flagellar shape for the sinusoidal initial condition up to $t=t''$ (white dashed line in Fig. \ref{fig5}b) at equal time intervals ($20$ ms). At $t=t'$, wave interference changes the direction of wave propagation. The full movie can be seen in Supplementary Movie 3. $\mbox{Sp}=5$, $\mu=100$, $\mu_a=2000$, $\eta=0.14$, $\zeta=0.4$, $L=50$ $\mu$m and $\tau_0=50$ ms. Arrows indicate the direction of wave propagation. }
\label{fig5}
\end{figure}

Finally, we study bending initiation and transient dynamics for two different initial conditions, in order to understand the selection of the unstable modes. In Fig. \ref{fig5}a,b the spatiotemporal transient dynamics are shown for the case of an initial eigenmode solution corresponding to the maximum eigenvalue (Fig. \ref{fig5}a) and an initial sine perturbation in $\phi$, with equal constant bound motor densities (Fig. \ref{fig5}b). In case (b) travelling waves initially propagate in both directions and interfere at $t=t'$ (Fig. \ref{fig5}b,d and Supplementary Movie 3). However, in the steady state both the eigenmode and sine cases reach the same steady state solution, despite the sinusoidal initial condition being a superposition of eigenmodes. This result provides a strong evidence that the fastest growing mode is the one that takes over and saturates in the steady state.  In Fig. \ref{fig5}c the transient dynamics for case (b) are shown for plus and minus-bound dynein populations close to the tail ($s=9/10$). 
Both populations decay exponentially with characteristic time $\bar{\tau}$ to $n_0$ and begin oscillating in anti-phase around this value, in a tug-of-war competition.

\section{\label{sec:7} Discussion} 
In this work, we presented a theoretical framework for planar axonemal beating by formulating a full set of nonlinear equations to test how flagellar amplitude and shape vary with dynein activity. We have shown how the nonlinear coupling of flagellar shape and dynein kinetics in a `sliding-controlled' model provides a novel mechanism for the saturation of unstable modes in flagellar beating. Our study advances understanding of the nonlinear nature of the axoneme, typically studied at the linear level \cite{Camalet, Camalet_NJP, Riedel-Kruse, Sartori:16}. \\

The origin of the bending wave instability can be understood as a consequence of the antagonistic action of dyneins competing along the flagellum \cite{Riedel-Kruse,Howard_Review}. The instability is then further stabilized by the nonlinear coupling between dynein activity and flagellum shape, without the need to invoke a nonlinear axonemal response  to account for the saturation of the unstable modes, in contrast to previous studies \cite{Bayly2014,Bayly2015}. Moreover, the governing equations (Eqs. \ref{Eq9_2} and \ref{Eq10}) contain all the nonlinearities in the limit of small curvature, and they are not the result of a power expansion to leading nonlinear order \cite{Hilfinger}.  Far from the Hopf bifurcation, linearized solutions fail to describe the flagellar shape and nonlinear effects arise in the system solely due to motor activity. At the nonlinear level, both the tangent angle and dynein population dynamics exhibit relaxation or cusp-like oscillations at some regions along the flagellum. Similar cusp-like shapes for the curvature have also been reported in sea urchin sperm \cite{Ohmuro}. This phenomenology is characteristic of motor assemblies working in far-from-equilibrium conditions and has been found in other biological systems such as in the spontaneous oscillations of myofibrils \cite{Julicher_spontaneous, Yasuda}.  Interestingly, despite the low duty ratio of axonemal dynein, a fraction of $\sim 2$ \% bound dyneins along the flagellum is sufficient to drive micrometer-sized amplitude oscillations. 

Angular deflections are found to be $\sim 0.1$ rad in the experimentally relevant activity range $\mu_a \simeq  (2-6) \cdot 10^{3}$ and the order of magnitude seems not to be crucially affected by viscosity nor other parameters in the system. Hence, despite of the fact that our description provides an intrinsic mechanism for amplitude saturation, it is only able to generate small deflections, typically an order of magnitude smaller than the ones reported for bull sperm flagella \cite{Riedel-Kruse}. Other structural constraints, such as line tension, are likely to influence the amplitude saturation, due to the elastohydrodynamic coupling with motor activity (see Supplementary Text). The presence of tension on the self-organized beating of flagella was previously investigated at leading nonlinear order; however, deflections were also found to be small \cite{Hilfinger}. Hence, we conclude that a `sliding-controlled' mechanism may not be sufficient to generate large deflections. Our work adds to other recent studies were the `sliding-controlled' hypothesis seems to lose support as the main mechanism responsible for flagellar beating \cite{Bayly2015, Sartori:16}. Basal dynamics and elasticity are also likely to influence the amplitude saturation, and substantial further research is still needed to infer whether sliding-controlled regulation is the responsible mechanism behind flagellar wave generation. \\

Principal component analysis allowed us to reduce the nonlinear dynamics of the flagellum in a two-dimensional shape space, regarding the flagellum as a single-frequency biological oscillator \cite{Ma}. Notice that a two-dimensional description would not hold for multifrequency oscillations, where an additional dimension is required \cite{Oriola_EPL}. Interestingly, we found that as activity increases, the characteristic wavenumber of the system slightly decreases. Thus, dynein activity has an opposite effect on wavenumber selection when compared with the medium viscosity (see Fig. \ref{fig2}d, inset). We also showed that the steady state amplitude is selected by the fastest growing mode under the influence of competing unstable modes, provided that the initial mode amplitudes are sufficiently small. \\ 

An important aspect which is not studied explicitly in this work is the direction of wave propagation. For simplicity, we used clamped boundary conditions at the head which are known to induce travelling waves which propagate from tip to base in the sliding-controlled model \cite{Camalet, Bayly2015}. It is beyond the scope of our study to assess the effects of different boundary conditions and the role of basal compliance at the head of the flagellum, which are known to crucially affect wave propagation \cite{Riedel-Kruse}. The present work also restricts to the case of small curvatures; however, the full nonlinear equations including the presence of tension could in principle be numerically solved as in previous studies \cite{Gadelha1, Hilfinger} (see Supplementary Text). Finally, real flagella is subject to chemical noise due to the stochastic binding and unbinding of dynein motors. Recent studies have provided insights on this problem by investigating a noisy oscillator driven by molecular motors. However, their approach was not spatially extended \cite{Ma}. Our approach could be suitably extended to include chemical noise in the system through Eq. \ref{Eq10} by considering a chemical Langevin equation for the bound dynein populations including multiplicative noise \cite{Gillespie}. In particular, it can be easily deduced from our study that by considering a force-independent unbinding rate, fluctuations of bound motors around the base state have mean $N \eta$ and variance $N \eta(1-\eta)$, in agreement with the results in Ref. \cite{Ma} where a different model was considered. \\

The possibility to experimentally probe the activity of dyneins inside the axoneme is one of the most exciting future challenges in the study of cilia and flagella. These studies will be of vital importance to validate mathematical models of axonemal beating and the underlying mechanisms coordinating dynein activity and flagellar beating. 

\section*{\bl{Ethics statement}}
{\bl Not relevant to our work.}

\section*{Data accessibility}
Electronic Supporting information for this article includes a Supplementary Text and 3 Supplementary Movies. {\bl The data from the Supplementary Movies is available from the Dryad Digital Repository: http://dx.doi.org/10.5061/dryad.qs65q }

\section*{Competing interests}

We have no competing interests.

\section*{Authors' contributions}

D.O. carried out analytical work, numerical simulations, data analysis and drafted the manuscript. H.G. and J.C. conceived of the study, coordinated the study and helped draft the manuscript. All authors gave final approval for publication.

\section*{\bl{Acknowledgements}}
{\bl No acknowledgements.}

\section*{Funding}
J.C. and D.O. acknowledge financial support from the Ministerio de Econom\'ia y Competitividad under projects FIS2010-21924-C02-02 and FIS2013-41144-P, and the Generalitat de Catalunya under projects 2009 SGR 14 and 2014 SGR 878. D.O. also acknowledges a FPU grant from the Spanish Government with award number AP-2010-2503 and an EMBO Short Term Fellowship with ASTF number 314-2014. H.G. acknowledges support by the Hooke Fellowship, University of Oxford. 

 \bibliographystyle{unsrt}

\end{document}